\documentclass[conference,compsoc]{IEEEtran}
\usepackage{url}

\makeatletter
\def\url@leostyle{%
  \@ifundefined{selectfont}{\def\UrlFont{\sf}}{\def\UrlFont{\small\ttfamily}}}
\makeatother

\ifCLASSOPTIONcompsoc
  \usepackage[nocompress]{cite}
\else
  \usepackage{cite}
\fi
\ifCLASSINFOpdf
  \usepackage[pdftex]{graphicx}
  \DeclareGraphicsExtensions{.pdf,.jpeg,.png}
\else
\fi
\usepackage{array}
\usepackage{url}

\begin{document}
%
%
\title{Towards In-Transit Analytics for Industry 4.0}
%
\author{\IEEEauthorblockN{Richard Hill and James Devitt}
\IEEEauthorblockA{School of Computing and Engineering\\
University of Huddersfield\\
United Kingdom\\
Email: \{r.hill\},\{j.devitt\}@hud.ac.uk}
\and
\IEEEauthorblockN{Ashiq Anjum and Muhammad Ali}
\IEEEauthorblockA{College of Engineering and Technology\\
University of Derby\\
United Kingdom\\
Email: \{a.anjum\},\{m.ali\}@derby.ac.uk}
}
%

%

%
%
\maketitle
%
\begin{abstract}
Industry 4.0, or Digital Manufacturing, is a vision of inter-connected services to facilitate innovation in the manufacturing sector. A fundamental requirement of innovation is the ability to be able to visualise manufacturing data, in order to discover new insight for increased competitive advantage. This article describes the enabling technologies that facilitate In-Transit Analytics, which is a necessary precursor for Industrial Internet of Things (IIoT) visualisation.\\
\end{abstract}
%
Industrial Internet of Things, edge computing, streaming analytics, data visualisation, digital manufacturing
%
%
%
\IEEEpeerreviewmaketitle
%
%
\section{Introduction}
%
%
The IoT is a logical progression as computing processing, storage and network infrastructure becomes both more accessible to use and cheaper to purchase. As manufacturers continue to reduce the physical form factor of wireless communication equipment, together with a desire to converge myriad communication protocols, it has become more feasible to enable sensors, actuators and other devices to connect to networks and exchange data from \emph{Machine to Machine} (M2M).

\begin{figure}[!t]
\centering
\includegraphics[width=3.45in]{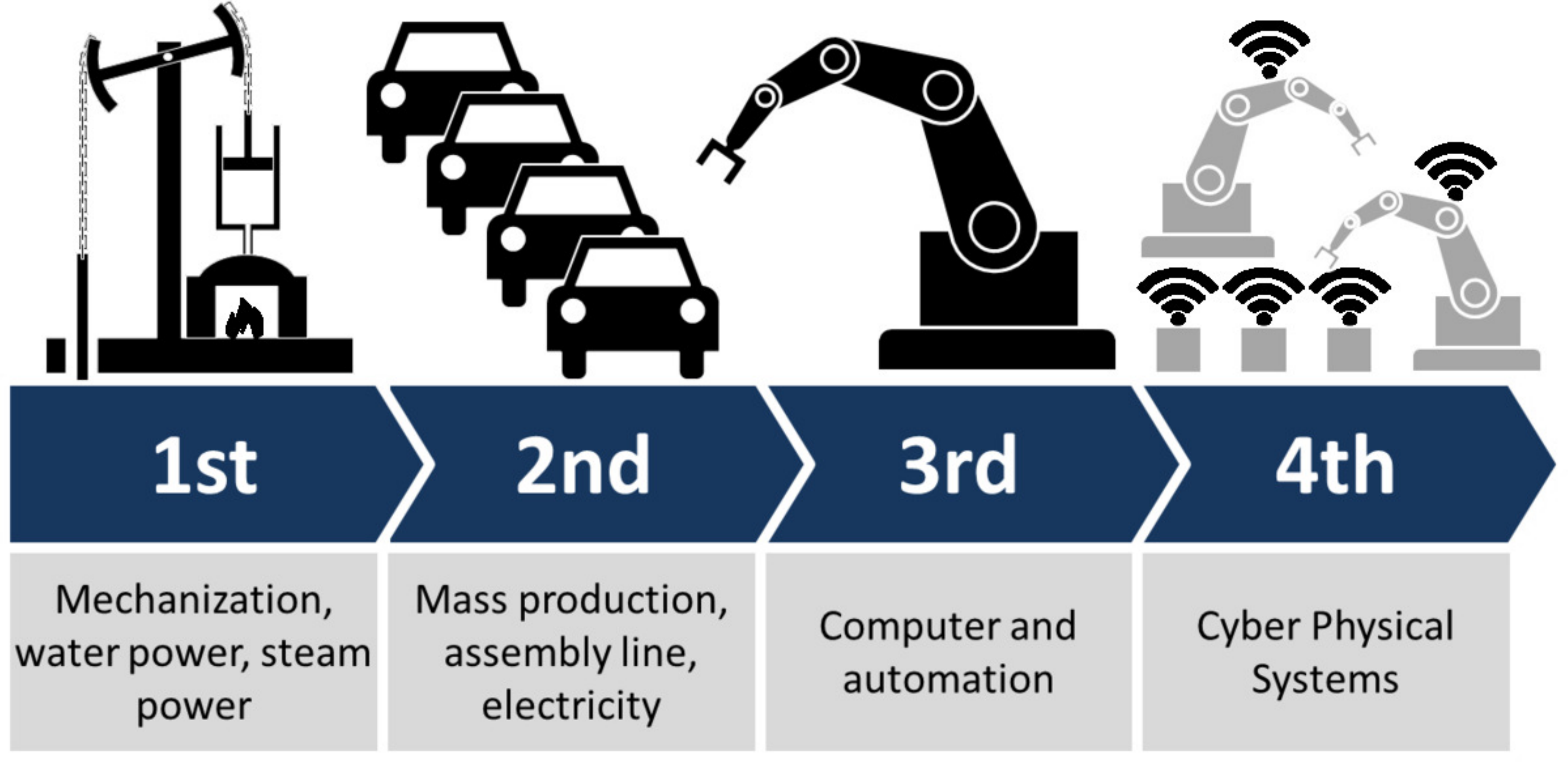}
\caption{Industrial revolutions and future view. By Christoph Roser, http://www.allaboutlean.com.}
\label{fig:revolutions}
\end{figure}

Significant events in the history of manufacturing have given rise to large step changes in the way that value is created. Fig \ref{fig:revolutions} illustrates the progression through various industrial revolutions. Perhaps most relevant to this discussion is the recognition of the third revolution, where computers and automation have become an intrinsic part of the manufacturing ecosystem, and most recently the declaration of the `fourth industrial revolution', where computers and automation become connected to form \emph{cyber physical systems}. Referred to as `Industry 4.0', the shift in focus is towards:
\begin{quote} ``the end-to-end digitisation of all physical assets and integration into digital ecosystems and value chain partners."\cite{pwc}
\end{quote}
As such the manufacturing industry has been an adopter of IoT technologies for some time, and the development of approaches to facilitate more open M2M communication is referred to as the \emph{Industrial Internet} or the \emph{Industrial Internet of Things (IIoT)}. Of the many definitions that exist, Ashton describes IoT as:
\begin{quote}
``a dynamic global network infrastructure with self-configuring capabilities based on standard and interoperable communication protocols where physical and virtual `Things' have identities, physical attributes and virtual personalities and use intelligent interfaces, and are seamlessly integrated into the information network''\cite{ashton}.
\end{quote}
%
%
%
More recently, discussions have arisen around the description of a manufacturing ecosystem that is wholly enabled by digital technologies, referred to as \emph{Digital Manufacturing}\cite{pwc}.

%

The successful digitisation of manufacturing value chains places a reliance upon the efficient, responsive and accurate exchange and analysis of data. This data is a vital part of feedback both to human observers and decision makers, as well as for the self-reliant autonomous control and coordination of manufacturing processes. As such, data, analytics and visualisation are inherent components of Industry 4.0.

%
\subsection{Organisation of article}
First, we review the key challenges faced by the manufacturing industry during its adoption of Industry 4.0.
We then propose In-Transit Analytics as an architectural approach to analysing streamed data in IIoT networks, followed by a case study in digital manufacturing, discussed in the context of the ITA architecture and a simulated set of results. Finally, conclusions will be drawn.
\section{Current challenges for manufacturing}
As mentioned above, manufacturing organisations demonstrate the most maturity at present in terms of the adoption of IoT architectures. Enterprises that make physical goods need to keep track and coordinate people, plant, tools and raw materials. The inter-connectedness of physical systems is not a new topic for the manufacturing industry, but it is only the recent efforts to miniaturise hardware, reduce costs of implementation, and to break-away from proprietary communication protocols that has accelerated the uptake of IoT approaches\cite{sundmaeker2010}.

\subsection{Data volume}
Wireless Sensor Networks (WSN) are an example of how IoT technology can be used to enable new possibilities\cite{bessis2013},\cite{li2015}.


Utility computing, such as cloud computing, has facilitated the growth of WSN in that scalable processing and storage can be requested on demand, without requiring a large investment in infrastructure and hardware\cite{liu2014}. Organisations that own large data centres make the necessary investment and provide the cloud computing utility on a pay-per-use basis.

If we now consider a collection of IoT devices as similar to the organisation of a WSN, we can see that alongside the potential to create new ways of working, there is also the prospect of producing vast quantities of data. This increase in data production creates significant challenges for its subsequent analysis, contextualisation and comprehension\cite{chen2015}.

\subsection{Visualising the value chain}
Although interconnected machines that exchange data between themselves and central IT systems is not a new concept for the manufacturing industry, such connectivity was often proprietory and ``siloed" in nature; data communication between machine tools might not be connected to subsequent production processes, therefore limiting the intelligence that could be available to guide or automate manufacturing system decisions. Since connectivity between the various organisational units within a manufacturing system is restricted, there is a risk of preventing the visualisation of a `complete picture' and in many cases this directly inhibits innovation.
IoT opens up manufacturing beyond process-centric visualisation towards a wider, sector-based view, that facilitates cooperative and collaborative, cloud-enabled business intelligence strategies\cite{hussain2012} across geographical regions, sectors of industry or both.

\subsection{Resource constraints}
The growth of devices becoming interconnected has been rapid, and will only increase at a greater rate\cite{gartner2015} as more devices become network enabled through either retrofit or as manufacturers include this capability into new devices. As IoT becomes more pervasive, increased volumes of data will be harvested, together with a greater demand for analytic capability within clouds for data visualisation. Therefore, both network bandwidth and computational power are constraints of current architectures for the further development of IoT. This is a major challenge for domains such as manufacturing, where in-process data is produced in large quantities\cite{shu2015}.

\subsection{Data velocity}
An additional challenge of more M2M communication is data \emph{velocity} - the rate at which data is produced - which is an area of particular interest to those in the Big Data research community. As we find more situations in which to embed sensors, or more machines to control, we also increase the rate at which data is produced. Process level data is produced in real-time, and as the potential value of process monitoring is realised, more real-time data will be collected by manufacturers. This will place more strain on the traditional bottleneck of the network infrastructure; the gateway between IoT devices and utility computing services for storage, processing and visualisation\cite{wang2016a,wang2016b}.

\subsection{IT architecture}
Traditional Information Technology (IT) environments, such as a client-server network architecture, might approach the design of networks with reference to a three, or \emph{n}-tier model, to distribute the processing and storage demands across a range of resources. As IoT devices have developed and matured, there is now more potential for compute power and storage to be located within or close to the device that senses or actuates; in effect the incorporation of processing and storage at the \emph{edge} of the network, commonly referred to as \emph{Edge} computing\cite{buyya2013}.

The enhancement of devices to include processing and storage means that varying degrees of analysis can be performed on continuous streams of real-time data, prior to eventual transfer of the data to a cloud-based data centre.

In the remainder of this article we propose an architecture that enables the exploitation of local compute power and storage at the edge of networks, to facilitate analytics processing upon streamed data from IoT devices.

\section{In-Transit Analytics}
%

Table \ref{tab:scenarios} illustrates three different scenarios of data use. Analysis of historical data is the most traditional understanding of data processing in an enterprise system. As the costs of computational power and storage have reduced, organisations have been able to process larger queries upon their repositories. Similarly, organisations have also utilised enterprise systems to forecast and plan the deployment of resources.
The use of data in the present has been less prevalent, and primarily has been focused upon the process under control. The computational requirements of data that is being continuously produced are intensive, as well as making great demands upon network infrastructure for any data that is transported away from the process.
\subsection{Edge computing}
Edge computing is a design approach whereby data processing capabilities are distributed across the whole of a network infrastructure. The widespread adoption of utility (cloud) computing has encouraged researchers and industry to explore how IoT devices can use their networking capabilities to interface directly to clouds; this arrangement allows many devices to connect to one central repository (a cloud), giving the benefits of centralised control and governance over data storage and repositories. 
There are however, some limitations associated with this approach as follows:
\begin{itemize}
\item A proliferation of devices presents additional complexity for network design, especially when there is a desire to introduce vast numbers of miniature wireless sensors and devices\cite{beer2003};
\item The cloud model of utility computing has drastically reduced computing infrastructure costs, but it is challenging to optimise the performance of centralised resources when they are required to maximise the conflicting demands of  storage and high power processing (in some cases HPC)\cite{ikram2015};
\item A central cloud connected to many individual network devices tends to assume that a common communication protocol will be employed, which is contrary to the real world manufacturing scenario of many different proprietary protocols and standards, each with different requirements depending upon the criticality of safety or quality in relation to the process being networked.
\end{itemize}
%
In the case of a manufacturing process that is producing data continuously, an edge computing architecture will make use of the local compute capabilities to process the data streams in real-time. 
\subsection{Streaming analytics}
Streaming analytics is an approach whereby data is processed continuously as it is produced, rather than retrospectively after it has been gathered and stored in a database. Analytics operations carried out on streaming data means that:
\begin{itemize}
\item Less data is transported across the network, which can bring an associated reduction in the amount of eventual data stored. This is evident when sensors that produce `noisy' data are filtered at source;
\item Visualisation functions can be sited closer to the origin of the data, which presents new opportunities for users to interact with the visualisations and augment/annotate them with additional data prior to being stored;
\item Central storage facilities can place an emphasis upon data mining, pattern recognition and predictive analytics, without the overhead of cleaning and conditioning raw data.
\end{itemize}
As a consequence of all of the above, knowledge can be stored as models (abstractions) of the raw data\cite{yaseen2017}, reducing the amount of redundant data that is retained in a repository, and increasing the quality and value of knowledge that is stored centrally.
\begin{table}[!t]
\renewcommand{\arraystretch}{1.3}
\caption{Data processing scenarios.}
\label{tab:scenarios}
\centering
\resizebox{8.6cm}{!}{
\begin{tabular}{|c|c|c|c|c|}
\hline
\textbf{Scenario} & \textbf{Function} & \textbf{Storage} & \textbf{Compute}   & \textbf{Network}                                                                 \\ \hline
Past              & \begin{tabular}[c]{@{}l@{}}Learn from\\ historical data.\end{tabular}                       & Large                                                     & \begin{tabular}[c]{@{}l@{}}Intensive but\\ infrequent.\end{tabular}                     & Low demand.                                                                      \\ \hline
Present           & \begin{tabular}[c]{@{}l@{}}Understand data\\ in relation to\\ current context.\end{tabular} & Small                                                     & \begin{tabular}[c]{@{}l@{}}Intensive.\\ Real-time data\\ at high velocity.\end{tabular} & \begin{tabular}[c]{@{}l@{}}Demanding.\\ High data transfer\\ rates.\end{tabular} \\ \hline
Future            & Predict, forecast.                                                                          & \begin{tabular}[c]{@{}l@{}}Medium\\ to large\end{tabular} & \begin{tabular}[c]{@{}l@{}}Intensive but\\ infrequent.\end{tabular}                     & Low demand.                                                                      \\ \hline
\end{tabular}
}
\end{table}
\section{Example scenario: automated inspection}\label{section:example}
\begin{figure}[!t]
\centering
\includegraphics[width=3.45in]{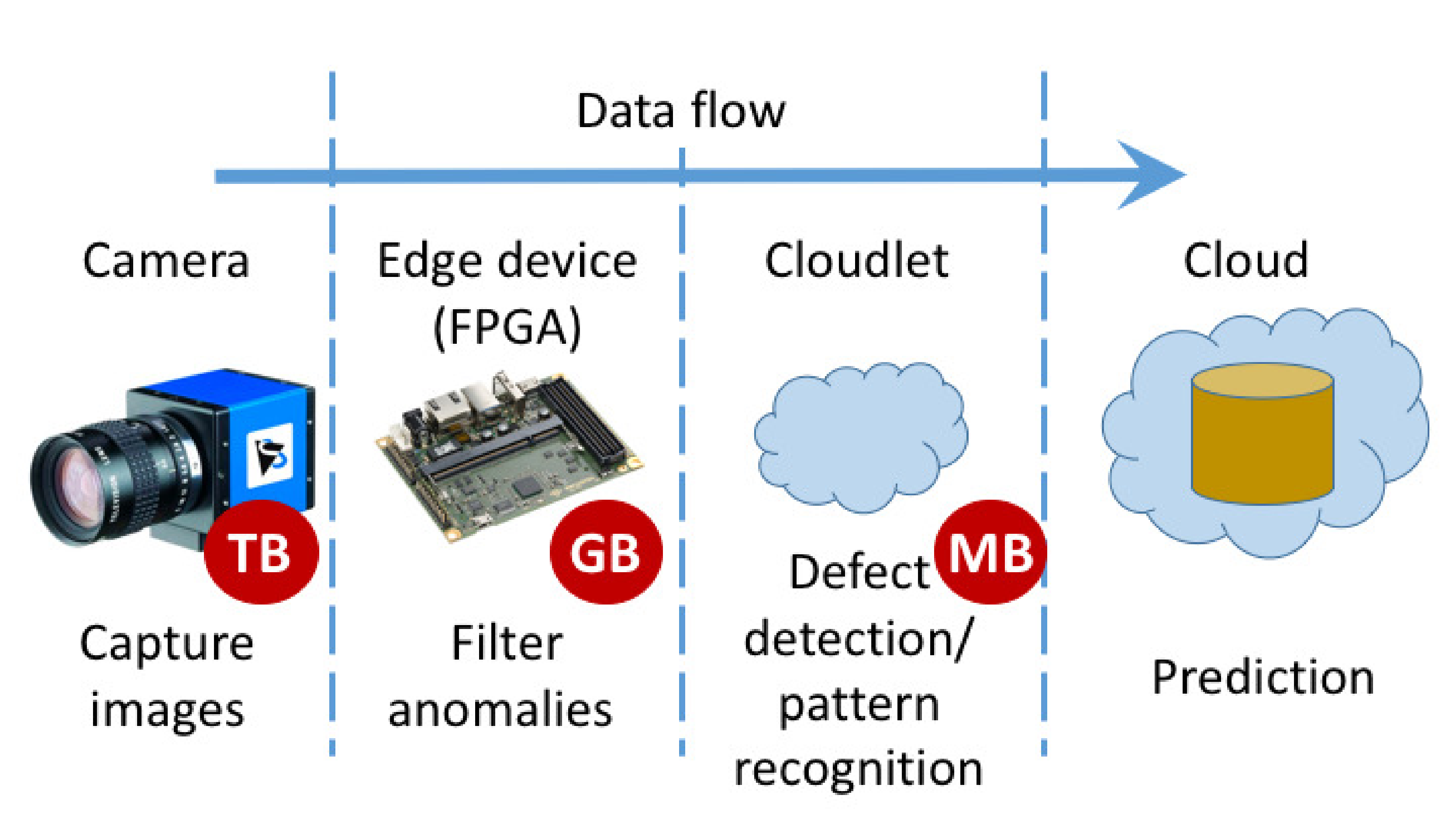}
\caption{In-Transit Analytics for defect detection process.}
\label{fig:itacasestudy}
\end{figure}

Using a particular example, we shall now explore an architecture that utilises In-Transit Analytics to facilitate improved connectivity across manufacturing value chains.
The inspection of defects in manufactured goods is an important part of the maintenance of quality standards from processes which can deliver variable outputs. Visual inspection is one example of quality control that can be used to identify products that need rework or repair, or which need to be rejected as not meeting a particular threshold standard.
%
%
Depending upon the extent of the visual inspection, it can be feasible to automate such processes using machine vision. In this example, the architecture for an automated inspection system that supports a move towards Industry 4.0 needs to take account of the following:
\begin{itemize}
\item Ensure that the real-time performance of OT equipment is maintained, alongside any specific safety requirements (SCADA). Such functionality should be kept separate from Ethernet/IT networking infrastructure, insofar as process integrity and safety is not dependent upon the IT  infrastructure. 
\item Maintain a desire to enable more intelligent M2M communication between devices, improving process control and communication locally at process level. 
\item Distribute analytics processing across a multitude of devices in order to reduce the transmission of data packets to the central IT infrastructure.
\item Enabling the central IT compute capability to process pre-conditioned data for increased insight into connected process-centric devices;
\item Facilitating more insightful visualisation closer to the data source, that can be a valuable source of interaction data from users.
\end{itemize}
The diagram in Fig \ref{fig:itacasestudy} illustrates how the ITA architecture influences the enablement of machine vision inspection for Industry 4.0. Images (either single frames or video) are captured by the inspection camera. A limited amount of processing is performed by the camera (image compression), before the image data is passed to a Field Programmable Gate Array (FPGA), which is an embedded systems device that can be re-programmed to suit different needs. An FPGA is an example of the type of hardware that is required to be SCADA compliant, yet it is now feasible to scale the computational power and storage in a way that was not previously possible with more traditional embedded systems.

The FPGA is located at the edge of the network and is used to process the camera data stream by filtering for any anomalies (potential defects). This requires processing power that is beyond that of the sensing device (the camera), and because of the inclusion of Graphics Processing Units, computationally intensive data conditioning can take place. This immediately reduces data feeds in the order of Terabytes down to Gigabytes after post-processing, which starts to alleviate the network of data that is redundant for the purposes of defect identification.
As an edge device, the FPGA is connected to a Cloudlet, which is a small-scale data centre also located at the edge of the network. Further processing can take place upon the output data streams, which have now been filtered and conditioned to some degree, ready for more detailed analysis, such as the application of pattern recognition techniques to identify a specific type of defect.

Finally, the output (which is now in the order of Megabytes) is available for storage within the central IT infrastructure such as a cloud. The cloud is then used as a repository for enterprise resource planning as well as for predictive analytics and modelling.
At each stage of the architecture there is an opportunity to visualise the data. For instance, filtration of anomalies at the FPGA edge device could be used by an operator so that they can aid the training of the system. The operator would examine the exceptions that are presented by the system and provide expert guidance to augment the data collected, assisting identification further downstream.
Similarly, visualisation at the cloudlet provides feedback for the control of the process. This can be observed at a local level, which is the traditional process-centric view that persists within the manufacturing industry.
%
%
%
FPGA hardware can already host software that utilises Machine Learning techniques, presenting more possibilities to improve the resolution, quality and usefulness of data that is streamed from edge devices.
\section{System Architecture}
\begin{figure}[!t]
\centering
\includegraphics[width=3.40in]{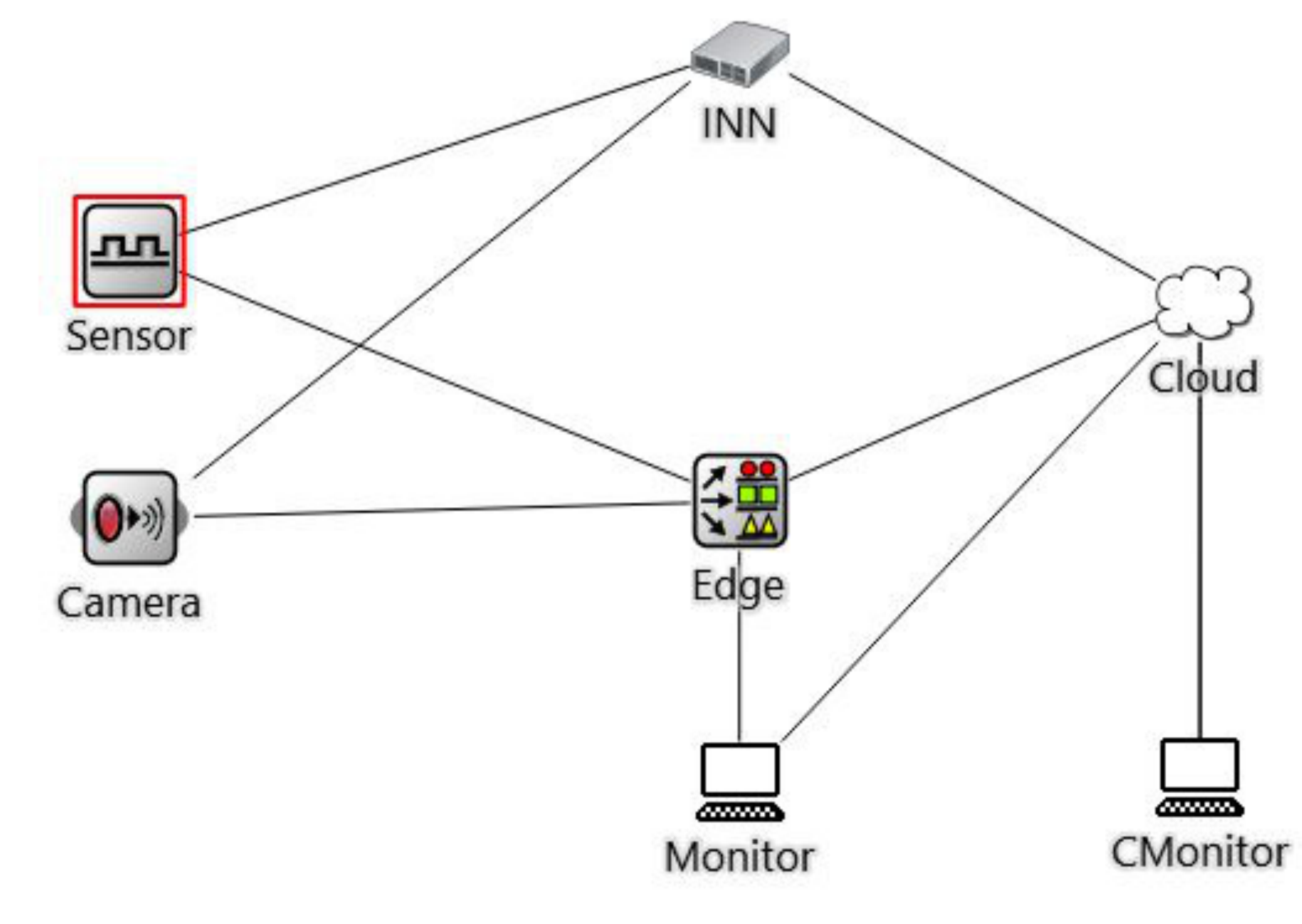}
\caption{Simulated In-Transit Analytics architecture.}
\label{fig:itasimulation}
\end{figure}

We have investigated the potential for improved performance that is possible with the ITA architecture. As per Fig \ref{fig:itacasestudy}, the ITA architecture consists of three tiers. The first tier represents data sources such as sensor and camera devices. Edge processing and storage capability such as FPGA/embedded systems and in-transit analytics processing nodes reside in the second tier of the ITA architecture. In the third tier, we represent cloud-based central processing and storage systems. It is assumed that components in the second tier will be located in closer proximity to first tier data sources, than the centralised resources of tier three. The architecture can be improved for different workflows to maximize compute and data optimization as discussed in \cite{habib2013}.
The simulation has been implemented in OMNet++ 5.1 (\url{https://omnetpp.org}). It consists of components, connections and messages which represent the traffic flow between the components.
Fig \ref{fig:itasimulation} illustrates the organisation of the model which consists of the following:
\begin{itemize}
\item \textbf{Sensor} - a component that provides data in relation to physical sensing of an environment e.g image sensor e.g camera or raw data sensor e.g temperature, humidity, light, etc.
\item \textbf{Camera} - a camera device which generates image/video data after regular intervals e.g 20 frames per second
\item \textbf{Intermediate Network Node (INN) } - a network node that enables data packets to be transported without performing any processing upon the message content e.g router, switch,hub.
\item \textbf{Edge} - an \emph{in-transit} component that has the capability to process and/or store limited amounts of data. This might be used for pre-processing, filtering/conditioning or transformation of data prior to its transfer to subsequent components.
\item \textbf{Monitor} - this component can be used to visualise data, as well as providing a means of input to interact with system parameters. For instance, we might modify the analysis criteria of an edge node component. 
\end{itemize}
The sensor component in the simulation has been configured to produce a temperature reading at certain intervals defined by the sensor delay parameter, which specifies the delay in seconds. For example, a value of `1.0' means that the sensor will be read every second. Similarly, a camera device produces image data at regular intervals defined by the camera delay parameter representing a real camera device monitoring some industry scenario and continuously producing data in the network.
As shown in Fig \ref{fig:itasimulation}, there are two data production sources (a sensor and a camera device) within Tier 1. Tier 2 contains \emph{INN} and \emph{Edge} components. Both \emph{INN} and \emph{Edge} receive the same copy of messages from the sensor and camera components through a wired connection. The lines in Fig \ref{fig:itasimulation} depicts two way wired connection between network resources,whose speed is defined by channel delay parameters as shown in Table \ref{table:channelparameters}.
\begin{table}
\centering
\caption{Connection Speed between different components}
\begin{tabular}{|l|c|c|}
\hline
\textbf{Connection} & \textbf{Speed} \\ 
\hline
Sensor and Camera to INN & 1 Gbps\\ \hline
Sensor and Camera to Edge & 1 Gbps \\ \hline
INN to cloud & 100 Mbps \\ \hline
Edge to Cloud & 100 Mbps \\
\hline \end{tabular}
\label{table:channelparameters}
\end{table}
%
%
%
\subsection{Analysis criteria}
A key challenge of a generalisable multi-tier and in-transit analytics architecture is to identify the criteria for analysis and algorithm processing required on each processing nodes of the network. The analytics requirements will vary depending upon specific applications and their domains.
Based on industry 4.0 scenarios, we have evaluated the in-transit analytics architecture by considering two specific cases. First, we look at an IoT scenario that makes use of simple sensors, producing raw data at regular intervals. Second, we study the more complex situation of camera device producing image data that may need post-processing.
%
%
In the first case, raw sensor data can be processed by a rule engine in near real time and is less CPU intensive, whereas image processing is a more CPU and data intensive task. Such processing has to be done under strict time constraints to provide feedback and real time analytics/visualisation. To assist with data intensive jobs, a meta-scheduling approach such as DIANA \cite{McClatchey2007} maybe used to significantly improve the execution time of the data intensive applications. For multiple clouds, scheduling algorithms such as \cite{Sotiriadis2015} can be used to improve the performance for number of metrics such as execution and turnaround time.
\subsubsection{Case One: Raw data processing}
\hfill \break
In raw sensor processing, the goal is to either forward or reject the message based on simple rule based processing. To process raw data, there are three stages:
\begin{enumerate}
\item Decode and determine the data priority.
\item Process the incoming message.
\item Decide whether to forward or reject the message.
\end{enumerate}
First, we decode the message and then determine its priority. In cases where streaming analytics cannot be achieved in near real time, it is necessary to prioritise the order in which data is processed. Processing of data is an application specific task, so understanding of the domain is necessary to devise the data priortisation order. Our experiments have utilised data provided by the MIT Artificial Intelligence Lab (\url{http://db.csail.mit.edu/labdata/labdata.html}), as the basis of the raw data processing scenario (Case One). It consists of data from 54 sensors with time, date, temperature, humidity, light and voltage readings. 
%
%
For sensor data which provides raw readings, the following rules apply:\\

\textit{Rule 1:For Sensor}\\
\begin{verbatim}
   if (temp>50)
   { {//forward data}
   else //do not forward   }
\end{verbatim}

\textit{Rule 2: For IoT device:}\\
\begin{verbatim}
   if (humidity<30 || light>35 ||
   voltage>500)
   {  { //forward data   }
      else //don't forward
   }  
\end{verbatim}
In both rules, we aim to forward data which is outside of a pre-defined range, indicating a significant event, which needs visualisation, monitoring or feedback from subsequent network nodes. In Rule 1, we only forward sensor readings where the temperature is greater than a threshold, in this case 50.
For Rule 2, we check for exceptional levels for the IoT device. Any data passing Rule 1 or Rule 2 will be forwarded to the next in-transit node.
%
%
%
\subsubsection{Case Two: Image based sensor processing}
\hfill \break
For Case Two, we use data coming from a camera device, which produces image data at a rate of 20 image frames per second. The goal is to analyse the data in real time, placing time constraints on the processing latency at the edge/in-transit nodes. 
As for Case One, the analysis consists of three stages. After decoding and data priority determination, we decompose our analytics algorithm into N stages, where N indicates the number of in-transit/edge and cloud nodes. We have two processing nodes, edge and cloud, and we divide our analytics algorithm into two parts, i) light part and ii) heavy part. For image datasets, we can perform image preprocessing, enhancement, dilation, scaling, transformation and similar (less CPU intensive algorithms) on the in-transit/edge nodes. On the cloud node, we can perform more CPU and memory demanding tasks such as image segmention, object detection, and deep learning algorithms.
To optimise edge processing, we define a parameter \emph{AnalyticsDeadline} which defines the time by which edge node can complete its tasks. The stage three forwarding or rejection rule for this case is given by: \\

\textbf{Algorithm for Image Analysis on the Edge node}\\
\begin{verbatim}
Mutex<DataBuffer> buffer;
  while(simulation){
   addIncomingMessageToBuffer();
   timer=startTimerThread();
  for each item in buffer{
   executeAlgorithm(item);
                         } }
TimerThread{
  if (timer.time>AnalyticsDeadlineTime)
   {
    sendBufferDataToCloud();
    //rejection case
    break;
   }}         
\end{verbatim}
The objective is to reduce the time taken by the cloud to perform complete analytics of the data under time constraints.
Performing part of the algorithm on the edge can also provide the end user with useful analytics/visualisation close to the data source as the data is \emph{in transit}. We have identified parameters which can be used to configure the behaviour of the edge node to conform with application based constraints in Table \ref{table:edge parameters}.
%

\begin{table}
\centering
\caption{Edge Node Analytics Parameters}
\begin{tabular}{|c|c|c|} \hline

\textbf{Name} & \textbf{Description} & \textbf{Value} \\ 
\hline
AnalyticsDeadline & Max Time for processing buffer &1 sec \\
\hline
BufferStorage & Edge items capacity & 20 \\
\hline
AlgorithmTime & Time to process one data item & 0.01 sec \\
\hline \end{tabular}
\label{table:edge parameters}
\end{table}
%
As shown in the pseudocode, \emph{AnalyticsDeadline} is implemented as a timer which is triggered after every \emph{x} seconds where \emph{x} is `1'. If an algorithm takes more than one second to perform the analytics, the time out is triggered, processing is halted and all the data items are forwarded to the cloud as they are. We seek to maximise the number of data items that are processed by the edge node before forwarding. By varying these parameters, system performance can be tested for different applications.
\subsection{Priority scheduling}
\label{priority}
We have adopted a workflow based approach to the consideration of how activity will be allocated to the architectural components. Data arriving from different sources (such as sensors and IoTs) is considered a \emph{job}. If several jobs arrive at an edge node for processing, the edge nodes may need to prioritise the processing of \emph{data type} (or value) first, followed by the \emph{source location}.
%
%
Based on prioritisation, a software defined network (SDN) and network function virtualisation (NFV) can be used to route data to different edge/in-transit nodes. With SDN and NFV, existing network infrastructure can be employed to provide in-transit computational capability under constraints to improve the efficiency of the analytics \cite{Zamani2017}.

\subsection{Results}
After determining the analysis criteria and the workflow task priorities for the dataset, the simulation produced a set of experimental results. Data produced from the sensor amounts to about 98 MB, and 990MB of image data for the camera, for the duration of the simulation (1000 seconds).  Data from both of these components was transported through edge and \emph{INN} components. The destination (cloud) component receives data from the \emph {Edge} and \emph{INN} components respectively.
\subsubsection{Case One: Raw Data Sensor Processing}
\hfill \break
\begin{figure}[!t]
\centering
\includegraphics[width=3.40in]{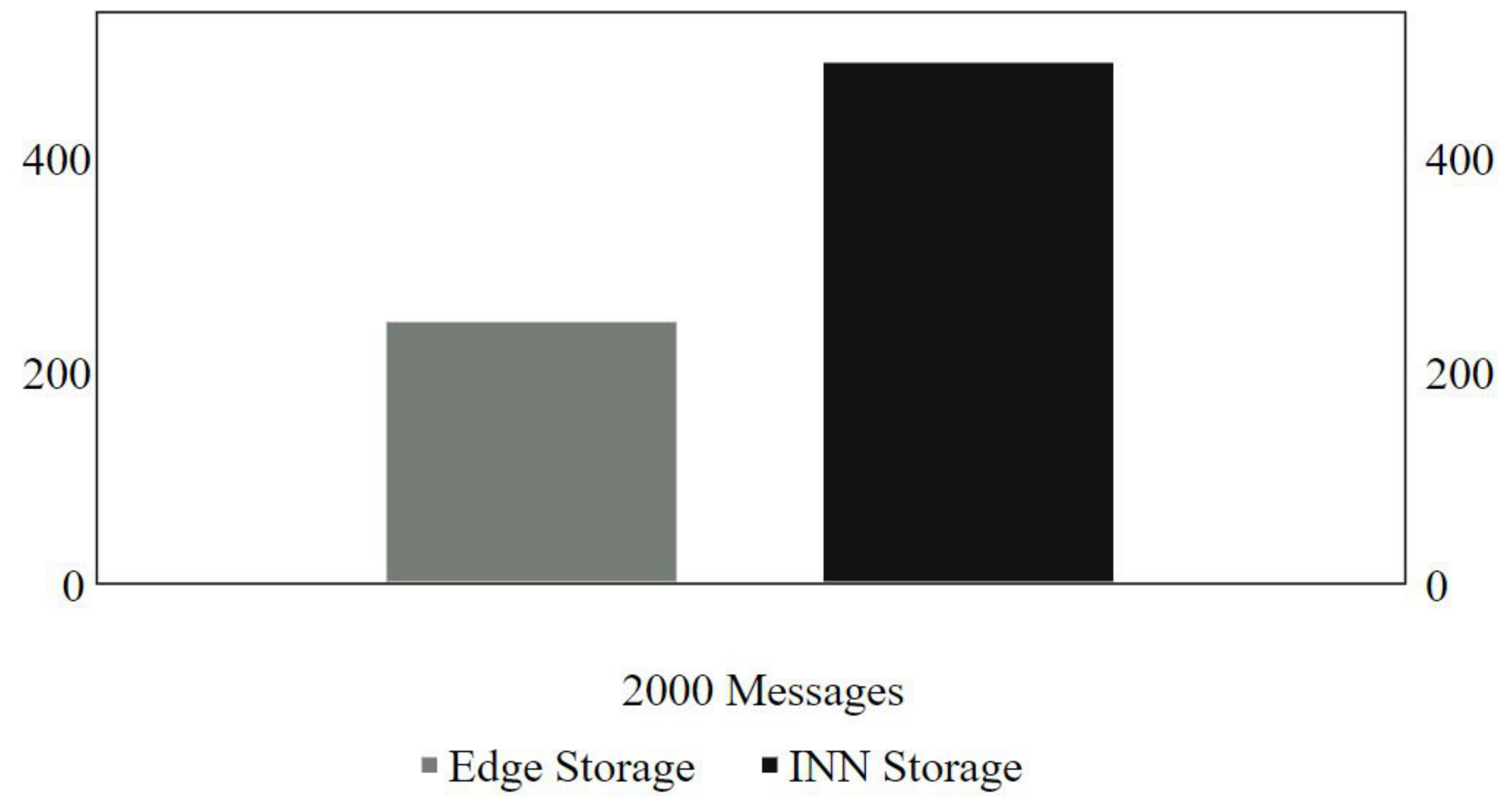}
\caption{Storage required by Edge and INN nodes on cloud.}
\label{fig:storage}
\end{figure}
As shown in Fig \ref{fig:storage}, data processing performed at the edge component has reduced the amount of data that was transferred to the cloud by 62\%, in comparison to the total data transferred via the \emph{INN} component to the cloud.
%
%
Eventually, all messages received at the destination cloud need to be processed, which consumes compute resources and time depending upon the individual algorithm or processing workflow. 

Within the simulation, the time taken by an algorithm to process one message is specified as parameter \emph{AlgorithmTime}, and can be used to calculate the total time taken by a cloud to process all of the messages. For example, If it takes 10 ms to process one message, by multiplying the individual message processing duration by the total number of messages to be processed, we can predict the total time taken by the cloud to process all of the messages.
%
\begin{figure}[!t]
\centering
\includegraphics[width=3.40in]{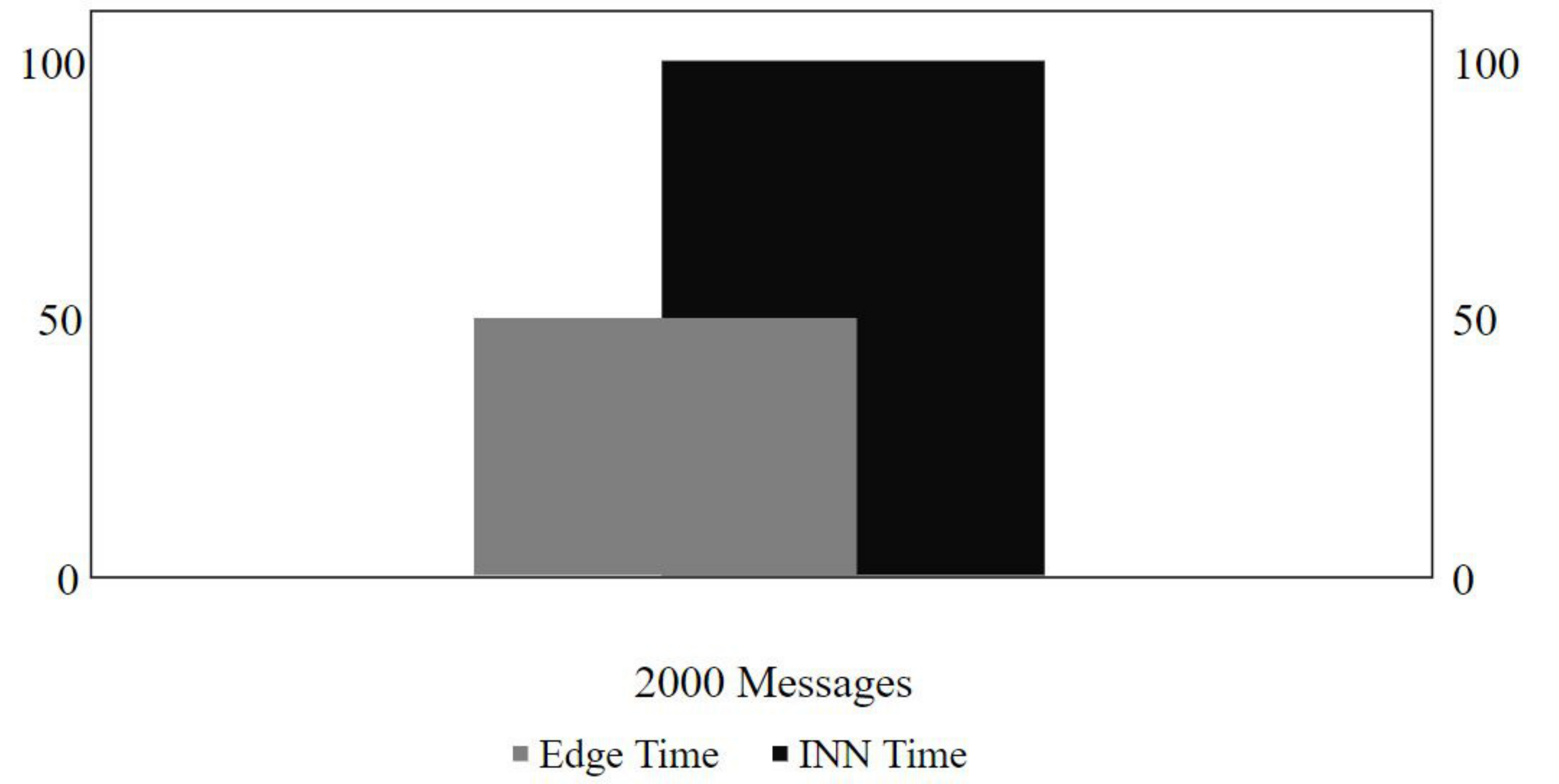}
\caption{Time taken to process edge and simple network messages.}
\label{fig:timetaken}
\end{figure}

Thus, as shown in Fig \ref{fig:timetaken} the time taken to process \emph{INN} messages is 20 seconds, and time taken to process edge messages is 9.3 seconds, a reduction of 10.7 seconds in total or 53.5\% saving in compute time.
\begin{figure}[!t]
\centering
\includegraphics[width=3.40in]{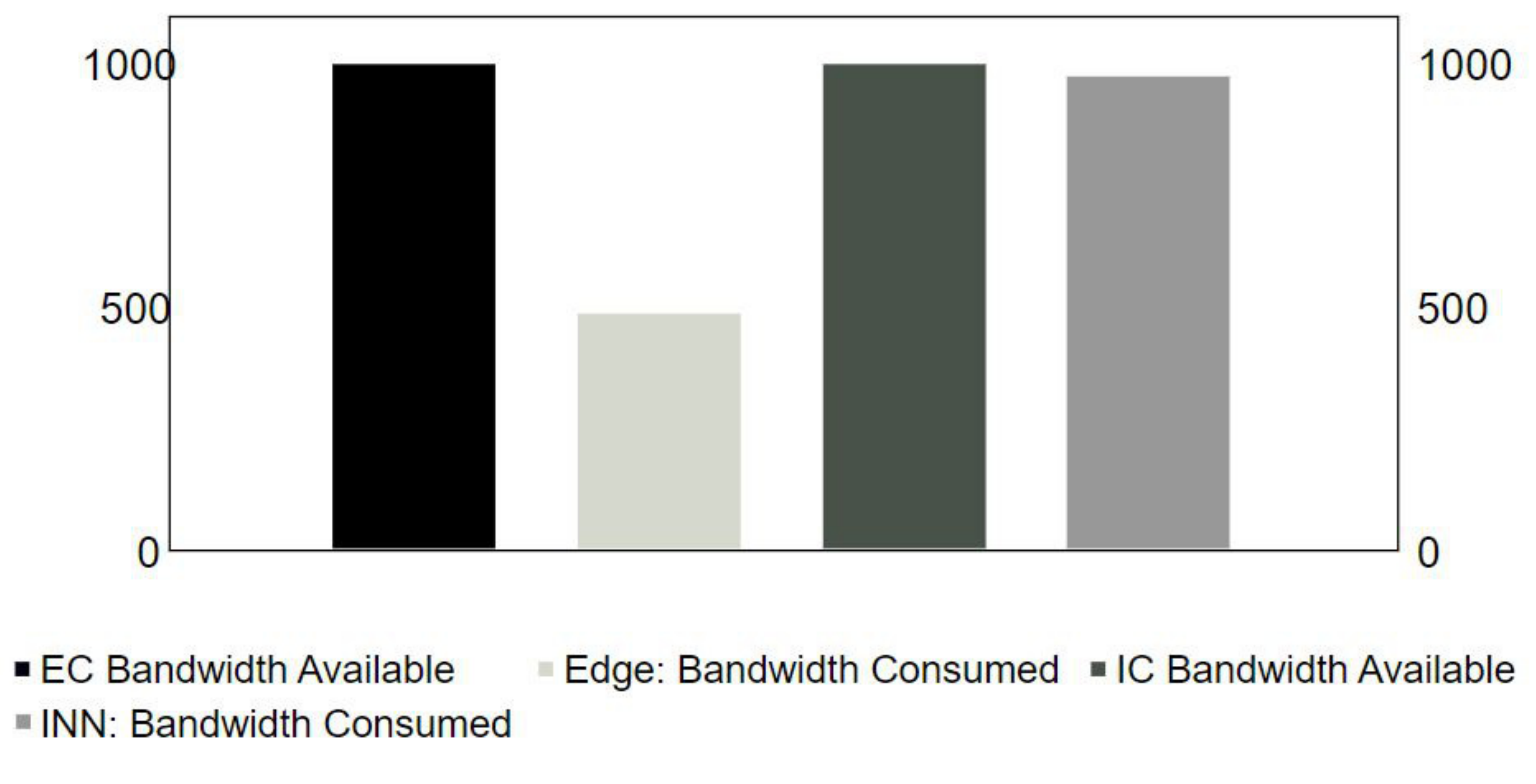}
\caption{Bandwidth consumed by Edge and INN messages.}
\label{fig:bandwidthConsumed}
\end{figure}
In Fig \ref{fig:bandwidthConsumed}, \emph{EC Bandwidth Available} indicates total connection bandwidth available from edge to cloud node and \emph{IC Bandwidth Available} shows the bandwidth available from INN node to cloud. As illustrated, bandwidth consumed by \emph{Edge node} is significally less than bandwidth consumed by the \emph{INN} (approximately 55\% less).

Whilst the same data was sent to both \emph{Edge node} and \emph{INN} nodes, due to edge processing, the amount of data was reduced and hence less bandwidth was ultimately consumed. It should also be noted that bandwidth consumed by \emph{INN} is approaching the total bandwidth available from \emph{INN} to \emph{cloud}, while bandwidth consumed by \emph{Edge node} is within the bandwidth available.
\subsubsection{Case Two: Image Sensor Based Processing}
\begin{figure}[!t]
\centering
\includegraphics[width=3.40in]{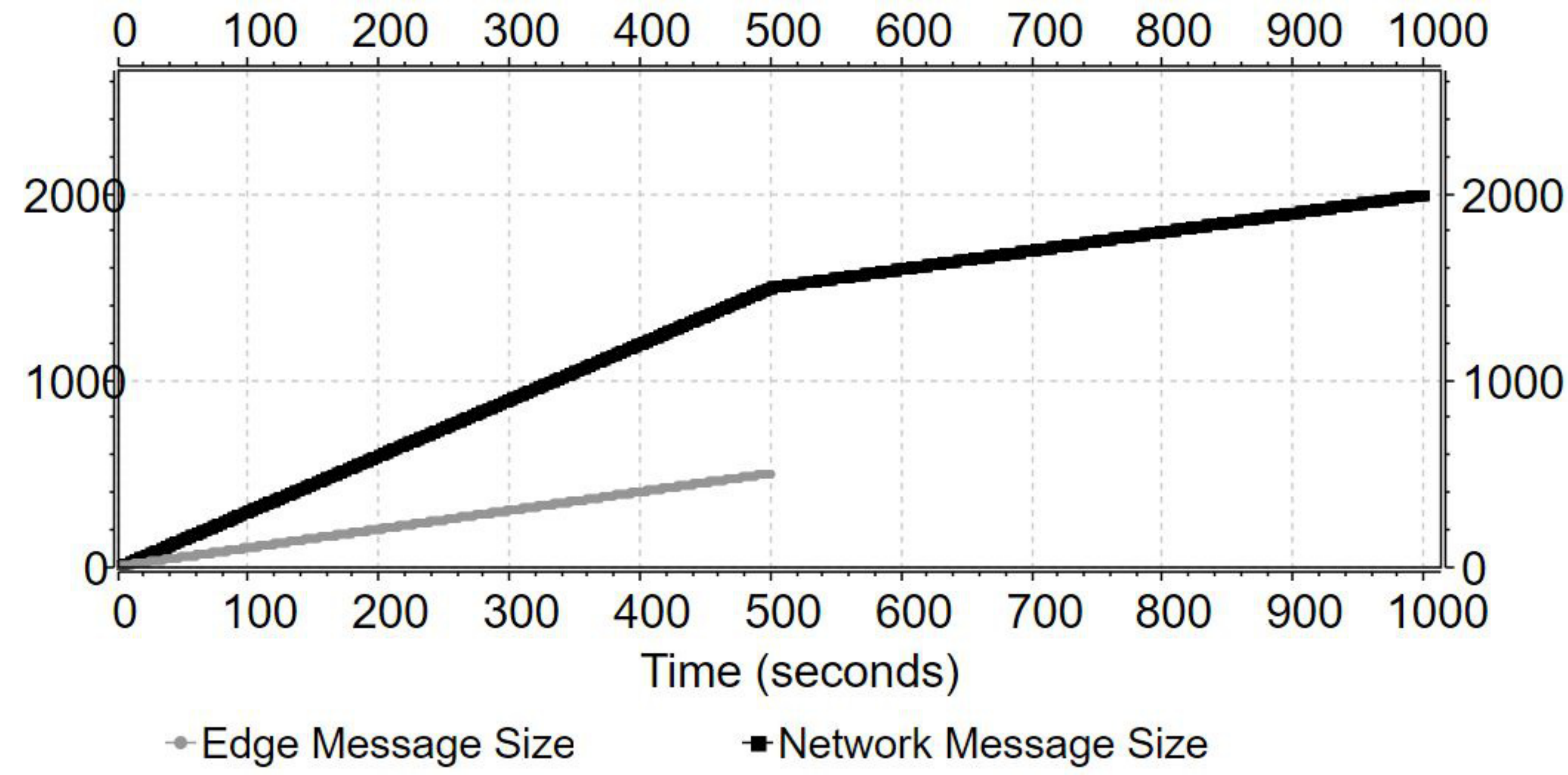}
\caption{Number of messages processed in time.}
\label{fig:processing}
\end{figure}

\hfill \break
In this case images from a camera sensor are computed under stringent time constraints. Messages which cannot be processed due to \emph{AnalyticsDeadline} timeout or buffer overflow are forwarded to the cloud without processing.
Fig \ref{fig:processing} shows the cloud processing of \emph{INN} and \emph{edge} messages. If messages are processed by an edge node under constraints e.g. buffer not full, and the \emph{AnalyticsDeadline} has not timed out, they are not forwarded to the cloud.

As shown in fig \ref{fig:processing}, for the first 10 seconds, the number of \emph{edge node} and \emph{INN node} messages processed by \emph{cloud node} are identical.  After 500 seconds, \emph{cloud node} has finished processing all of the \emph{edge} messages, as the sensor data rate is twice that of the camera device and also the number of \emph{edge node} messages to process are less than the \emph{INN} messages (about 600 in contrast to 2000 \emph{cloud} messages).

The \emph{cloud} continues to process the \emph{INN} messages until the end of the simulation (t=1000). In conclusion, for INN node, \emph{cloud node} has to process 2000 messages, and in case of \emph{edge node} with in-transit processing, \emph{cloud node} only has to process 600 messages making the edge based network more productive by approximately 70\% in this case.

\begin{figure}[!t]
\centering
\includegraphics[width=3.40in]{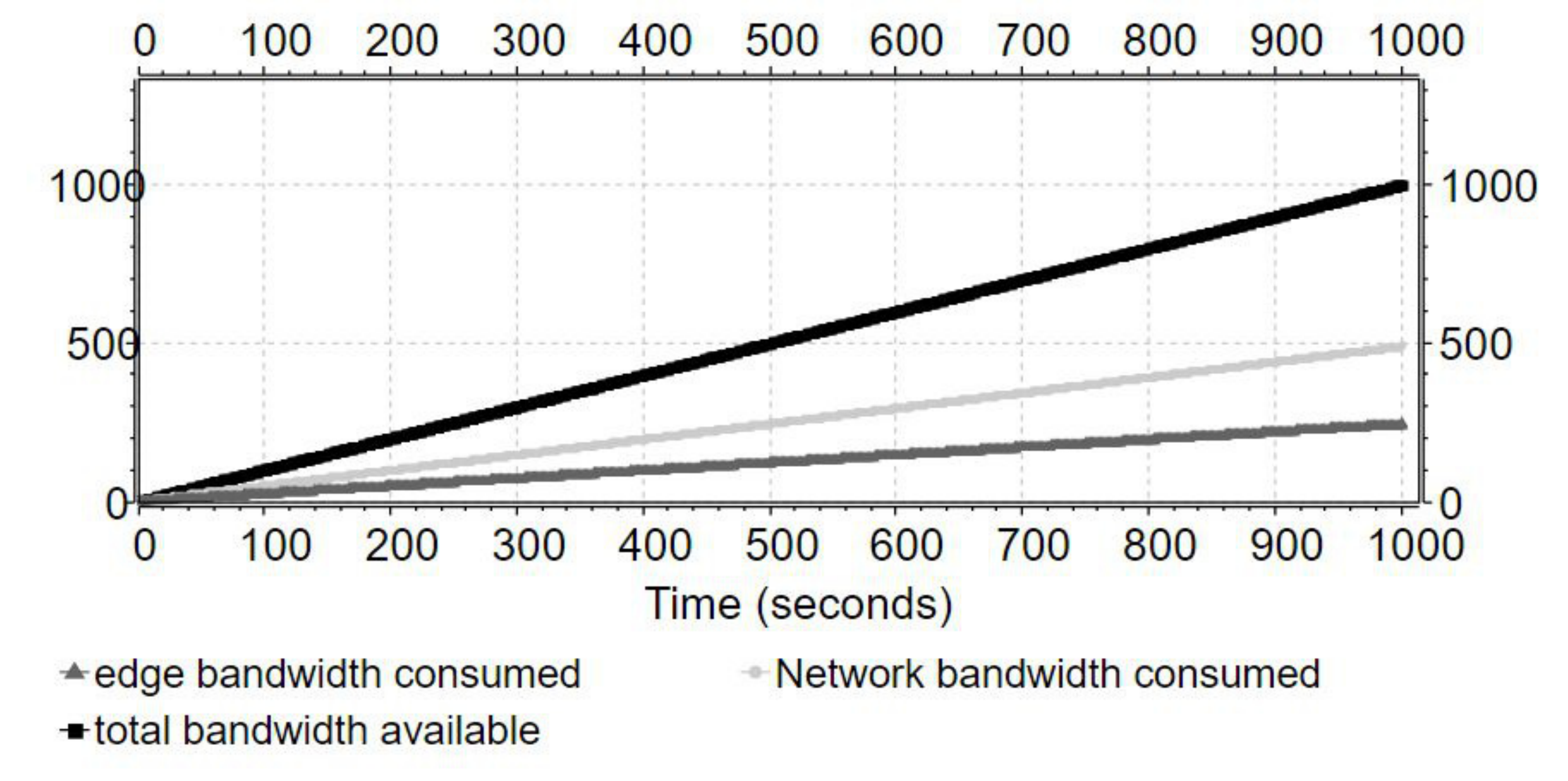}

\caption{Bandwidth consumed with one sensor and IoT device.}
\label{fig:bcn}
\end{figure}

The efficiency gained by incorporating edge computing depends on a range of factors and parameters setting for e.g analytics criteria, algorithm time, number of edge nodes and physical topology of the network including communication medium and channel speeds.
Fig \ref{fig:bcn} shows the bandwidth consumed by \emph{edge node} and \emph{INN} nodes with respect to time as the messages are in-transit. The total bandwidth for connections from \emph{Edge} to \emph{Cloud} and \emph{INN} to \emph{Cloud} is approximately 2GB. The network bandwidth consumed by \emph{INN} shows processing of all the messages during simulation interval of 1000.
The graph is generated with one sensor and one camera device and indicates a steady-state network simulation, as bandwidth required by both \emph{edge node} and \emph{INN} components is less than the total bandwidth of the channel. A key aspect of these results is the reduced bandwidth required by the Edge node (56\% less), as compared to an \emph{INN}.
\begin{figure}[!t]
\centering
\includegraphics[width=3.40in]{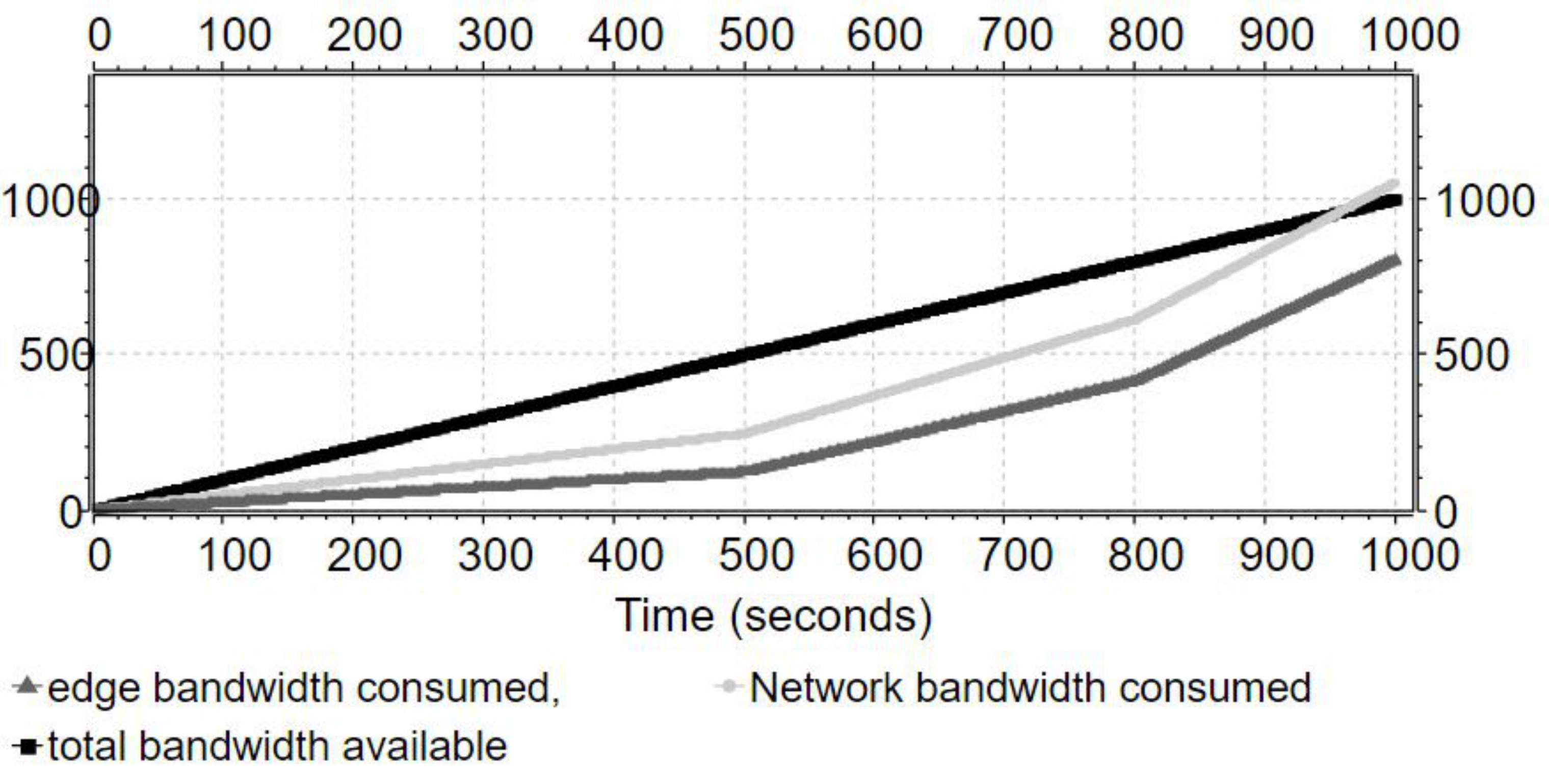}
\caption{Bandwidth consumed with 1, 4 and 8 sensors.}
\label{fig:bca}
\end{figure}
We again calculate the bandwidth consumed in fig \ref{fig:bca}. However, in addition to sensor and camera devices, after t=500, we add 3 other sensor nodes to observe the effect of the bandwidth consumption from \emph{Edge} and \emph{INN} nodes.

After t=800 we add another 4 sensors to determine the effect  on the bandwidth consumed. As mentioned earlier both \emph{Edge} and \emph{INN} receive redundant copy of the messages from sensor and camera device and are independent from each other. The simulation is run for 1000 seconds, from 0 to 500 simulation seconds and the graph is identical to Fig \ref{fig:bcn}. However, after t=500 (when 3 extra sensors start producing), data rates increase and more bandwidth is required by both \emph{Edge} and \emph{INN} nodes.

We observe that the curve for \emph{Edge} bandwidth consumed is less steep than the curve for INN, which indicates that the \emph{Edge} node is less prone to changes in network than the \emph{INN} node. At t=800 with an additional sensors are added there is a steep rise in the data rate consumed by \emph{INN} node and reduced rise by the edge node. At t=900, just before the simulation is about to finish, the data rate capacity required by \emph{INN} exceeds the total bandwidth available on the network, whereas data rate capacity required by the Edge node is still below the total bandwidth curve. This graph proves, edge computing can reduce the bandwidth requirement of the network topology, in this case saving approximately 54.\%. 

\section{Conclusions}
We have considered some of the issues facing the manufacturing value chain whilst it undergoes the fourth industrial revolution. Specifically, we have identified data analytics and visualisation as fundamental topics that can assist manufacturers to gain the capability to be truly digital; this requires the interconnection of devices (through IoT technologies) as well as innovative approaches to include myriad legacy architectures and equipment, to enable the free exchange of dat from M2M.

Our proposed architecture, In-Transit Analytics, serves to augment existing legacy SCADA and IT architectures with the capability to capture and process data at different stages of the data pipeline. The use of an edge analytics approach enables significant data transfer savings whilst also facilitating richer, more relevant analytics to occur, whilst also mitigating the endless progression of capturing and storing more data, some of which is often redundant.

\end{document}